\documentclass[10pt,twocolumn,letterpaper]{article}

\usepackage{cvpr}
\usepackage{times}
\usepackage{epsfig}
\usepackage{graphicx}
\usepackage{amsmath}
\usepackage{amssymb}
\usepackage{comment}
\usepackage{caption}
\usepackage{url}
\usepackage{color, soul} 

\usepackage[breaklinks=true,bookmarks=false]{hyperref}
\hypersetup{
    colorlinks=true,
    linkcolor=blue,
    filecolor=magenta,      
    urlcolor=cyan,
}

\cvprfinalcopy 

\def\cvprPaperID{6084} 

\begin{document}

\title{Adversarial Video Compression Guided by Soft Edge Detection}

\author{Sungsoo Kim, Jin Soo Park, Christos G. Bampis, Jaeseong Lee,\\ Mia K. Markey, Alexandros G. Dimakis, Alan C. Bovik  \\
The University of Texas at Austin\\
Austin, Texas  78709, USA\\
{\tt\small (sungsookim, js.park, bampis, jason.lee27, mia.markey)@utexas.edu} \\  
{\tt\small
(dimakis, bovik)@ece.utexas.edu}
}


\twocolumn[{%
\maketitle
\renewcommand\twocolumn[1][]{#1}%
\begin{center}
\includegraphics[scale=0.36]{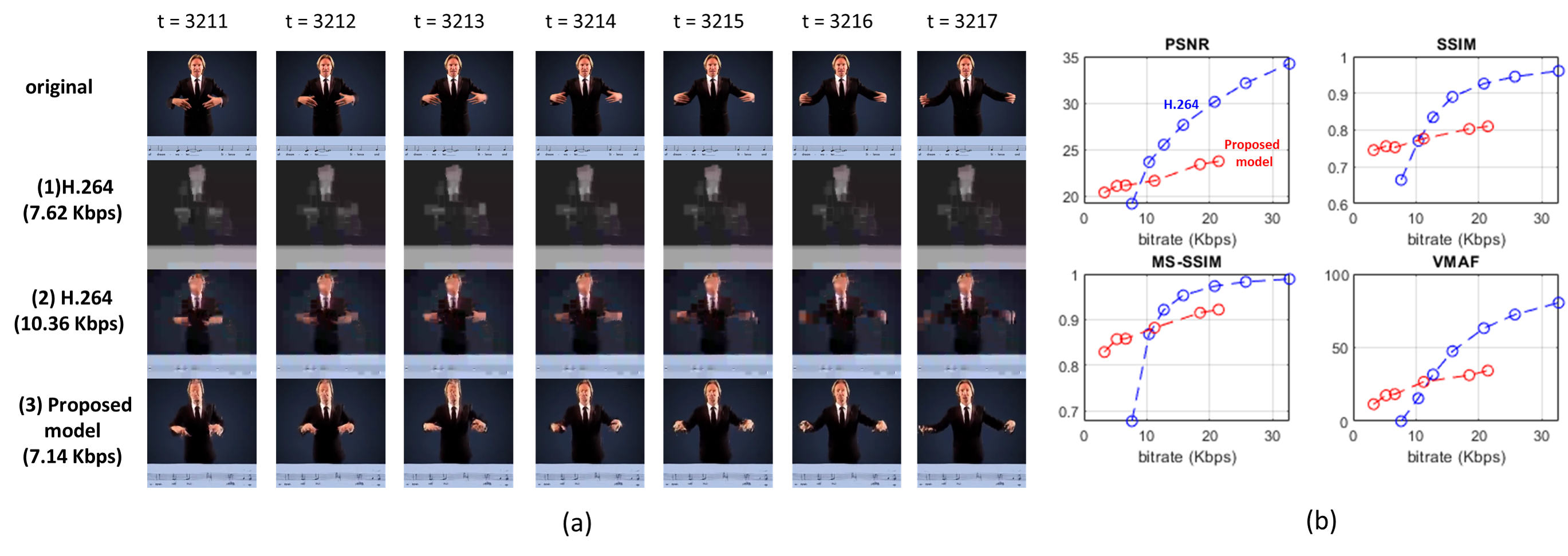}
   \captionof{figure}{(a) Seven consecutive frames from an original video~\cite{Charles16} 
   and three compressed videos, (1) H.264 at 7.62 Kbps, (2) H.264 at 10.36 Kbps, and (3) our proposed soft edge-guided GAN-based video compression at 7.14 Kbps. Our model delivered the best quality reconstructions at low bitrates. (b) RD-curves using four popular perceptual video quality assessment metrics. Our model (red curves) achieved much higher quality scores below 10 Kbps 
   compared to H.264 (blue curves). At bitrates below 7.5 Kbps, our model produced viable reconstructions while H.264 failed (no output). Please see
   \href{https://youtu.be/VHk7G5V6iBs}{\url{https://youtu.be/VHk7G5V6iBs}} for experimental results on several videos.}
\label{fig:signiture}
\end{center}%
}]

\begin{abstract}
We propose a video compression framework using conditional Generative Adversarial Networks (GANs). We rely on two encoders: one that deploys a standard video codec and another which generates low-level maps via a pipeline of \textit{down-sampling}, a newly devised \textit{soft edge detector}, and a novel \textit{lossless compression scheme}. For decoding, we use a standard video decoder as well as a neural network based one, which is trained using a conditional GAN. Recent ``deep" approaches to video compression require multiple videos to pre-train generative networks to conduct interpolation. In contrast to this prior work, our scheme trains a generative decoder on pairs of a very limited number of key frames taken from a single video and corresponding low-level maps. The trained decoder produces reconstructed frames relying on a guidance of low-level maps, without any interpolation. Experiments on a diverse set of 131 videos demonstrate that our proposed GAN-based compression engine achieves much higher quality reconstructions at very low bitrates than prevailing standard codecs such as H.264 or HEVC.
\end{abstract}

\section{Introduction}

Video compression is a process of reducing the size of an image or video file by exploiting spatial and temporal redundancies within an image or video frame and across multiple video frames. The ultimate goal of a successful video compression system is to reduce data volume while retaining the perceptual quality of the decompressed data. Standard video compression techniques, such as H.264, HEVC, VP9 and others, have aimed for decades to push the limits of these ``rate-distortion" tradeoffs. Developing better video compression techniques lies at the core of applications where more efficient video storage and transmission is essential, such as video streaming.

Despite the success of traditional video compression standards, there has recently been heightened interest in the possibility of neural network based image and video compression, fueled by the success of deep neural networks (DNNs) in image-related applications. DNNs have been successfully applied to image compression \cite{theis2017lossy, balle2016end} and shown to
deliver promising performance, especially at low bitrates~\cite{agustsson2018generative}. Similar ideas have been applied to video compression, e.g., by casting the motion estimation task as an interpolation solved by training on a large volume of videos~\cite{santurkar2018generative, wu2018video, wen2018generating}. These approaches have achieved performance approaching that of prevailing standardized codecs such as H.264 and HEVC~\cite{wu2018video}.  

Here, we propose a novel video compression framework that uses conditional Generative Adversarial Networks (GANs). Our proposed model automatically generates low-level semantic label maps using a newly conceived \textit{soft edge detector} combined with a \textit{down-sampler} (encoder). A conditional GAN-based network decoder is then trained on ``soft" edge maps and on specified key frames of a video. Only a relatively small number of key frames (1\%) of a single video is required, without the need for an interpolation process trained on multiple videos as in prior schemes~\cite{santurkar2018generative, wu2018video, wen2018generating}.

We compare our designed video codec to two widely-deployed state-of-the-art video compression standards (H.264 and HEVC) on two public video datasets; KTH~\cite{schuldt2004recognizing} and the YouTube pose dataset~\cite{Charles16}. A total of 215K frames (131 videos) are used in the comparisons. All of the data is publicly available, and our system will be released upon acceptance. 

\begin{figure*}
\begin{center}
\includegraphics[scale=0.48]{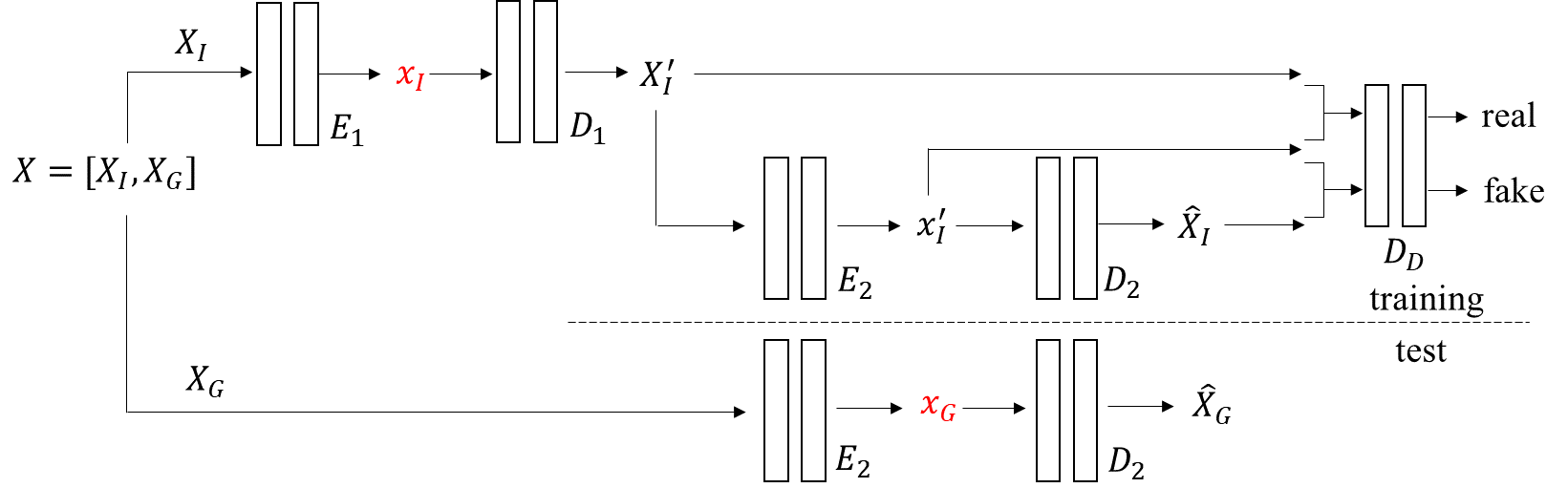}
\end{center}
   \caption{
   Proposed framework for adversarial video compression. Note that $X$ consists of only one video to be compressed. The video $X$ is partitioned into two sets containing different types of frames: $X_I$ and $X_G$. $X_I$ is lightly compressed into $x_I$ using the standard H.264 encoder, while $X_G$ is highly compressed into $x_G$,
   that contains only soft edge information at low resolution. The $X_I$ are used to train a generative model that we call the second-stage decoder $D_2$. This generative model is trained at the receiver using $x'_I$ and $X'_I$ using a discriminator $D_D$. After training, $D_2$ takes soft edges $x_G$ as input and produces reconstructed frames (see also Figure \ref{fig:downsampler_ex}). Only $x_I$ and
   $x_G$ are required to reconstruct the decompressed video.} 
\label{fig:framework}
\end{figure*}

\section{Related work}

\subsection{Semantic synthesis via adversarial learning}

Rather than using random latent codes in generative adversarial networks (GANs)~\cite{goodfellow2014generative}, conditional GANs include both deterministic and probabilistic traits of latent codes~\cite{mirza2014conditional}, to build a semantic relationship between a latent code and an output code. This approach has been used, for example, to translate pictures from one image space to another, given pairs of images as training data~\cite{isola2017image}. The idea of learning semantic relationships between two image spaces via adversarial training has been studied for other applications as well ~\cite{dong2017semantic,kaneko2017generative,karacan2016learning}.

\subsection{Generative adversarial networks for video}

More recently, GANs have been used for video applications, e.g., for predicting future frames; for learning video representations using Long Short-Term Memory (LSTM)~\cite{srivastava2015unsupervised}, for video prediction and generation~\cite{mathieu2015deep, vondrick2016generating, lotter2016deep, oh2015action}, for quantizing image patches into a dictionary~\cite{ ranzato2014video}, for estimating pixel motion~\cite{ finn2016unsupervised, liu2017video}, for spatio-temporal autoencoding~\cite{patraucean2015spatio}, and for stochastic video prediction~\cite{lee2018stochastic, denton2018stochastic} and generation~\cite{ liang2017dual, xiong2018learning}. GANs have also been applied to video compression by learning to generate interpolated frames using a large number of videos~\cite{ santurkar2018generative, wu2018video, wen2018generating}. While this approach has been effective, it is likely inappropriate for live video streaming, since long term frame predictions can produce severe and systemic artifacts (such as missing or blurred objects). Furthermore, the predictive accuracy of theses approaches can be degraded severely in the presence of large or sudden movements of objects. Our approach seeks to combat theses problems using soft edge-guided conditional GANs, as explained in Section~\ref{sec:VC_SED_cGAN}.

\subsection{Soft edge detection}

The detection of substantive changes in luminance, or edges, once a cornerstone of computer vision theory, is 
also regarded a plausible front-end feature extraction process in biological vision systems~\cite{marr1982vision}. While the importance of edge detection has somewhat faded, the zero crossings of bandpass derivative responses remain popular in low-level image analysis tasks~\cite{marr1980theory, canny1986computational}. 

One intriguing aspect of bandpass zero crossings is the plausibility of reconstructing a filtered signal from the zero crossings of multiple responses, motivated by Logan's original theorem~\cite{logan1977information}. While a number of approaches to image reconstruction from 2D bandpass zero-crossing maps have been proposed~\cite{yuille1985fingerprints, mallat1991zero}, the topic has found little currency in recent years, owing to the difficulty and sensitivity to noise and perturbations of the reconstructions. Nevertheless, because of the highly compact, binary nature of zero-crossing maps, the prospect remains intriguing, particularly in view of the possibility of training deep learning networks to learn zero-crossing based reconstructions on real data, with resilience against realistic noise, blur, and other distortions. 

In our approach, the frame representations are learned by a conditional GAN that is guided by soft edge responses. We adapt the terminology of ``soft" since our edge detector generates multilevel edge maps, rather than binary ones. We have found that the use of soft edges, rather than strictly defined binary hard constraints on the reconstruction, produces much better results (see Section~\ref{sec:exp_Q_edge}).
One important implication of our work is on re-establishing the importance of edges, as inspired by biological vision, but for informing a modern deep video compression architecture. 

\section{Preliminaries}

Let $X \in \mathbb{R}^{W \times H \times 3 \times N}$ denote the set of whole frames in a video, having a spatial resolution of $W \times H$ pixels, three ({\textit{RGB}}) channels and temporal duration of $N$ frames. Also, let $X^{(t)} \in \mathbb{R}^{W \times H \times 3}$ denote the $t$-th frame of $X$ and $n(\cdot)$ denote the cardinality of a set of frames, i.e., $n(X)=N$. Each channel in a video frame is stored using 8 bits. 

We start by partitioning the set of frames $X$ of the video into two subsets: $X_I$ and $X_G$. $X_I$ is the set of selected key frames and $X_G$ is the set of generated (non-key) frames, which we also refer to as ``G-frames". Let $n(X_I)=N_I$ and $n(X_G)=N-N_I$, where $N_I \in \{1,2,...N\}$. We can write:

\begin{equation}
\begin{split}
X_I&=\left\{{X_I}^{(1)}, {X_I}^{(2)}, ... ,{X_I}^{(N_I)}\right\} \\
X_G&=\left\{{X_G}^{(1)}, {X_G}^{(2)}, ... ,{X_G}^{(N-N_I)}\right\}.
\end{split}
\end{equation}
\noindent where $X_I$ and $X_G$ are composed of {\bf{any}} $N_I$ and $N-N_I$ frames from $X$, respectively. The elements of $X_I$ play a similar role as that of I-frames in conventional codecs (and convey similar advantages).

\section{Video Compression}
\subsection{Proposed framework}
\label{sec:VC_SED_cGAN}

Our GAN-based compression model has two encoding stages (see Figure \ref{fig:framework}). At the first stage, the frames in $X_I$ are compressed by the encoder $E_1$ to generate a representation $x_I$. Then, the decoder $D_1$ decodes $x_I$ to reconstruct $X_I'$. If $n(X_I)<N$, then $X_I'$ and $X_G$ further follow a second stage of encoding; the encoder $E_2$ encodes $X_I'$ and $(X_G)$ to generate $x_I'$ and $x_G$. Then, the second stage decoder ($D2$), decodes $x_I'$ and $x_G$, yielding the reconstructions $\hat{X}_I$ and $(\hat{X}_G)$. Notably, we can use any conventional codec to implement $E_1/ D_1$. Here, we implement $E_1/D_1$ using a conventional H.264 encoder (decoder), which is efficient and performs very well for key frame compression. If every frame is selected as a key frame, then our scheme degenerates to a frame-based H.264 encoder, without any motion prediction/compensation.

\begin{figure}
\begin{center}
\includegraphics[scale=0.4]{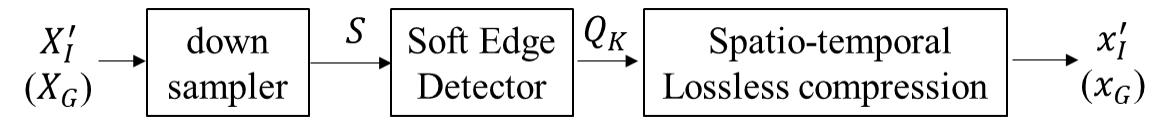}
\end{center}
   \caption{Overview of the second encoding stage ($E_2$).}
\label{figure:E_2}
\end{figure}

The second encoding stage is significantly different and involves a series of steps: \textit{down-sampling}, \textit{soft edge detection} and a novel spatio-temporal edge map compression scheme (see Figure \ref{figure:E_2}). The corresponding decoder $D_2$ is trained using a GAN and a discriminator $D_D$.

Naturally, the encoder $E_2$ cannot learn evolving representations of non-key frames using information from key frames only. Hence, we employ a conditional GAN to train $D_2$ using pairs of key frames and their corresponding soft edge maps. During training, $D_2$ learns associations between the key frames and the soft edge maps. Once $D_2$ is trained, it can be guided by the soft edge maps $x_G$ to reconstruct the G-frames $X_G$ (non-key frames), which the decoder has never seen before. When the soft edge map contains more information, it can guide the decoder to reconstruct frames with better quality.




\subsection{Second stage encoder, $E_2$}

The second encoding stage is composed of a \textit{downsampler} and a \textit{soft edge detector} which feeds a \textit{lossless compressor} (see Figure \ref{figure:E_2}). Let $S$ and $Q_k$ denote the outputs of the \textit{downsampler} and the \textit{soft edge detector}, and $x_I'$ and $x_G$ be the outputs of the \textit{lossless compressor} respectively. The downsampling step reduces the spatial resolution of each frame, but does not affect the number of frames. The purpose of downsampling is to produce a more compact representation that is easier to compress.


\begin{figure}
\begin{center}
\includegraphics[scale=0.39]{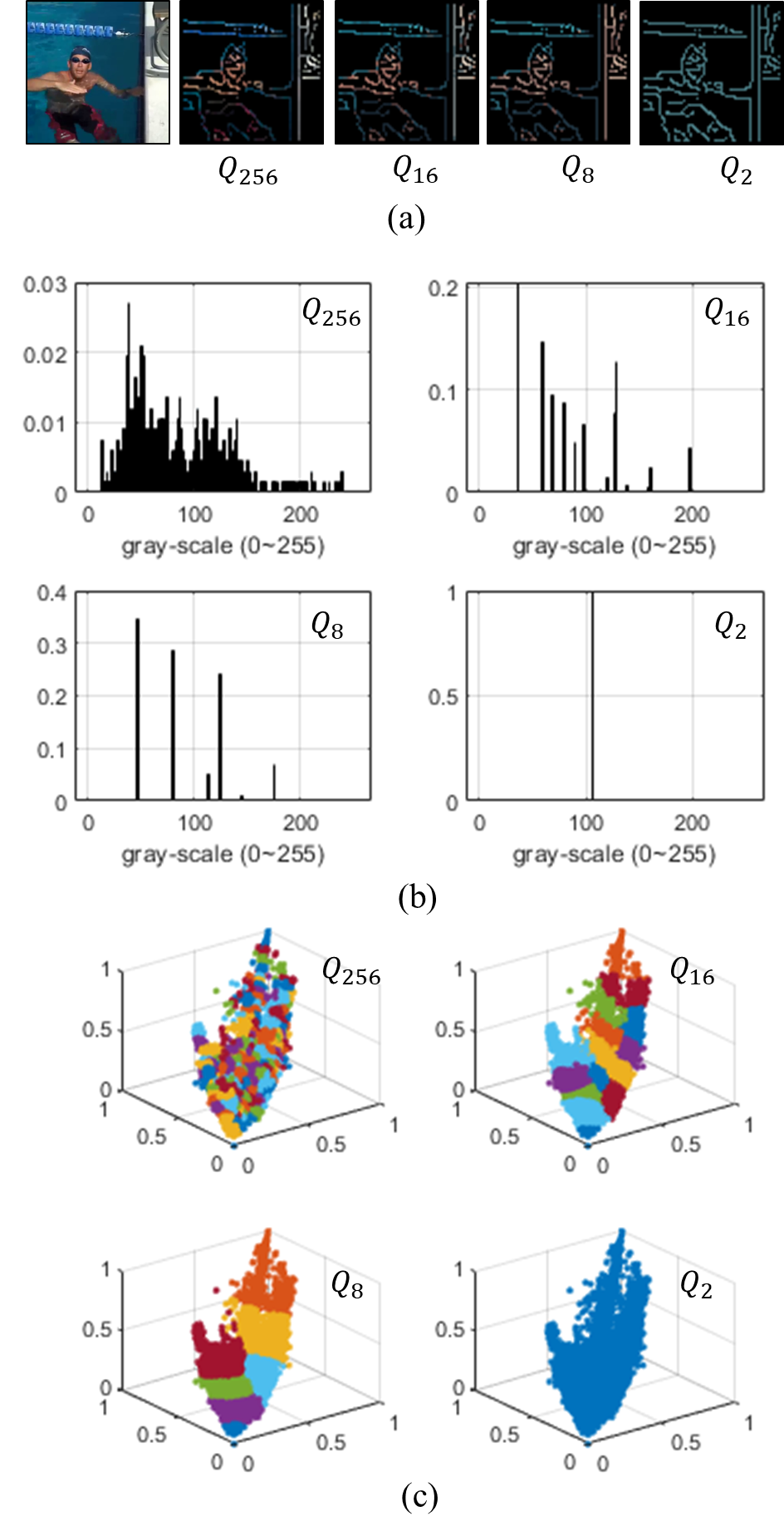}
\end{center}
   \caption{Outputs of \textit{soft edge detector}. (a) The left-most frame is a $64\times 64$ downsampled frame $S^{(1)}$ from a reconstructed frame ${X_I}^{(1)'}$ of one video~\cite{Charles16}. The right four frames are outputs of the \textit{soft edge detector} for different levels of quantization $k$ ($Q_k$). (b) Grayscale histograms of $Q_k$. (c) Three dimensional scatter plots (normalized R/G/B axes) of $S$, where colors visually distinguish the clusters indexed by $Q_k$.}
\label{figure:soft_edge_histogram}
\end{figure}

\subsubsection{\textit{Soft edge detector}}
\label{sec:Soft_edge_detector}

Following downsampling, we perform a soft edge detection step, which further produces a compressible representation of $S$. First, we apply the Canny edge detector~\cite{canny1986computational, marr1980theory} to find an edge pixel map $I = [I_{i,j}]$, where

$$
I_{i,j}=
\begin{cases}
    1,  & \text{if $(i,j)$ is an edge}\\
    0,  & \text{else},
\end{cases}
$$


Following edge detection, we perform vector quantization to form clusters of colored pixels mapped to edge pixels, i.e.,

\begin{equation}
q_{i,j}'=\mathbb{V}_k \left( s_{i,j} \cdot I_{i.j} \right)
\label{eq:soft_edge}
\end{equation}
\noindent where $s_{(i,j)}$ and $q_{(i,j)}$ are the $(i,j)$-th elements of the downsampled frame $S$ and a subsequent quantized frame $Q$. $\mathbb{V}_k (\cdot)$ is a vector quantizer which uses the $k$-nearest mean~\cite{cover1967nearest}, to form $k-1$ clusters of colored pixels  mapped to $I_{i,j}=1$. The large cluster of pixels $I_{i,j}=0$ is not included.



The aforementioned soft edge detection approach is illustrated in Figure \ref{figure:soft_edge_histogram}. In Figure \ref{figure:soft_edge_histogram} (a), we show a $64\times 64$ frame downsampled from a reconstructed $256\times 256$ frame
${X'_I}^{(1)}$
from one video of the {\textit{YouTube Pose dataset}}~\cite{Charles16}. 
The next four frames in Figure \ref{figure:soft_edge_histogram} (a) are outputs from the \textit{soft edge detector} for several different levels of quantization. As the quantization level $k$ is decreased, the cardinality of colors co-located with edges decreases. 
Figure \ref{figure:soft_edge_histogram} (b) plots the histograms of the gray-scale brightness in each $Q_k$.
Each video will have a different set of centroids for a given $k$. Figure \ref{figure:soft_edge_histogram} (c) depicts three-dimensional scatter-plots (R/G/B axis) of $S$ with different colors assigned to each of its clusters $Q_k$. The maps $Q_{256}$ and $Q_8$ in Figure \ref{figure:soft_edge_histogram} (a) appear quite similar, yet the entropy of $Q_{8}$ is much reduced relative to that of $Q_{256}$ (2.0904 vs 7.0358).

\begin{figure}
\begin{center}
\includegraphics[scale=0.35]{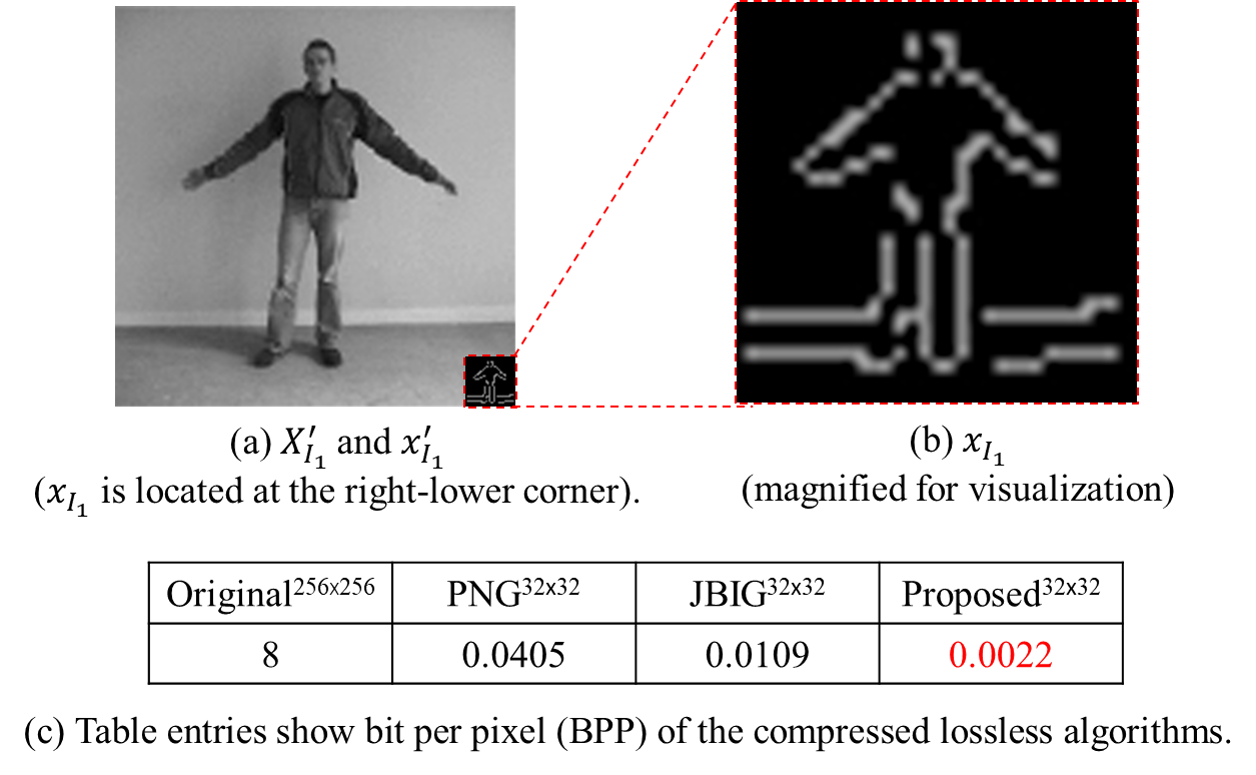}
\end{center}
   \caption{Efficiency in bits per pixel (BPP) achieved by different \textit{lossless compression} schemes on a bi-level image.}
\label{fig:lossless}
\end{figure}

\subsubsection{Lossless compression}

Since most of the outputs $Q$ of the \textit{soft edge detector} are zero, we found it useful to perform \textit{lossless} compression on the generated soft edge maps. However, and to the best of our knowledge, there have been no lossless compression schemes developed for soft edge maps. Therefore, we devised the following lossless compression scheme. We first applied run-length encoding~\cite{robinson1967results} on the soft edge maps along both the spatial and the temporal directions. Then, we performed another lossless compression step using Huffmann encoding~\cite{huffman1952method}.

Thanks to the sparse nature of the soft edge maps, our scheme produces great compression results compared to JBIG2~\cite{JBIG22011}, a state-of-the-art approach for bi-level image compression. As shown in Figure \ref{fig:lossless}), we achieve a $4.95\times$ compression gain over JBIG2, across all $567$ frames for a given test video.





\subsection{Second stage decoder, $D_2$}

We utilize the conditional GAN framework of Isola \etal~\cite{isola2017image} to train the decompressor $D_2$ and discriminator $D_D$ by parsing the compressed data $x_i'$ into both original data $X_I'$ and decompressed data $\hat{X}_I$ in an adversarial manner. Note that only key frames are used to train $D_2$ and $D_D$ (Figure~\ref{fig:framework}). The objective function of the conditional GAN can be expressed as
\begin{equation}
\begin{split}
L\left( D_2, D_{D} \right)
=\mathbb{E}_{X_I}[h(X_I)].
\end{split}
\end{equation}
Please refer to the supplementary material\footnote{Supplementary materials: \href{https://drive.google.com/file/d/1ojKDhTVq7CiyvAkVryfE3BUl1rvp2Ynq/view?usp=sharing}{\url{https://drive.google.com/file/d/1ojKDhTVq7CiyvAkVryfE3BUl1rvp2Ynq/view?usp=sharing}}} for the definition of $h(\cdot)$. $D_2$ and $D_D$ are trained on many pairs of frames generated from $X_I$.

\section{Experimental results}

\subsection{Evaluation and Datasets}

\textbf{Evaluation measure}: To conduct a quantitative performance analysis, several video quality assessment (VQA) metrics were deployed; PSNR, SSIM~\cite{wang2004image}, MS-SSIM~\cite{wang2003multiscale} and VMAF~\cite{NeflixVMAF}. We used the following two publicly available video datasets for our experiments:

\textbf{KTH dataset}~\cite{schuldt2004recognizing}:
The KTH dataset includes 600 short videos (6 types of actions, 25 people, 4 different scenarios). Each video contains between 400 and 800 frames. From these, we randomly collected 100 videos.

\textbf{YouTube Pose dataset}~\cite{Charles16}: This dataset is comprised of 45 YouTube videos. Each video contains between 2,000 and 20,000 frames, covering a broad range of activities by a single person: dancing, stand-up comedy, sports and so on. We excluded 13 of the videos owing to their long durations ($>$15 min; 7 videos), frequent scene changes (6 videos), and one duplicated video. The remaining 31 videos were included in our analysis. 

\subsection{Comparison to conventional schemes}

We compared the performance of our GAN-based compresssor with the currently prevailing codec H.264 (refer to the supplementary material for a comparison with H.265). We do not compare our scheme with recent video compression models based on DNNs~\cite{santurkar2018generative, wu2018video, wen2018generating}, since (1) our model is only trained on a subset of the frames from the single target video without any pre-training process and (2) our model does not rely on any interpolation. Our innovations are complementary to the other DNN-based compression schemes and can be fruitfully combined in future work. 

\subsection{Implementation}
~\textbf{Architecture of} {\boldmath{$D_2$}}: The second stage of the decoder $D_2$ is trained by a discriminator $D_D$. In our experiments, the original frames were re-sized and cropped to size $2^8\times2^8$ over eight consequent layers. A stride size of $2\times2$ was used when training the DNNs. Our model can be implemented using any options for resizing and cropping\footnote{We built an additional DNNs module that the number of layers and stride sizes could be designed to process original sized frames without resizing. For example, if an original video was of resolution $1280\times720$, then the DNN structure was configured automatically, with seven layers and seven sets of strides $(2,2), (2,2), (2,2), (2,3), (2,3), (2,5)$ and $(2,5)$.}. The batch size and the number of epochs were fixed at 1 and 1000, respectively. 

~\textbf{Architecture of} {\boldmath{$E_2$}}: The second stage encoder $E_2$ has a predetermined structure composed of a \textit{downsampler}, the \textit{soft edge detector} and the \textit{lossless compressor} (Figure \ref{figure:E_2}). To simplify experiments, the \textit{downsampler} only scaled the frames by factors for $4 \times 4$ and $8 \times 8$.

\subsection{Quality of reconstructed downsampled frame}
\begin{figure}
\begin{center}
\includegraphics[scale=0.25]{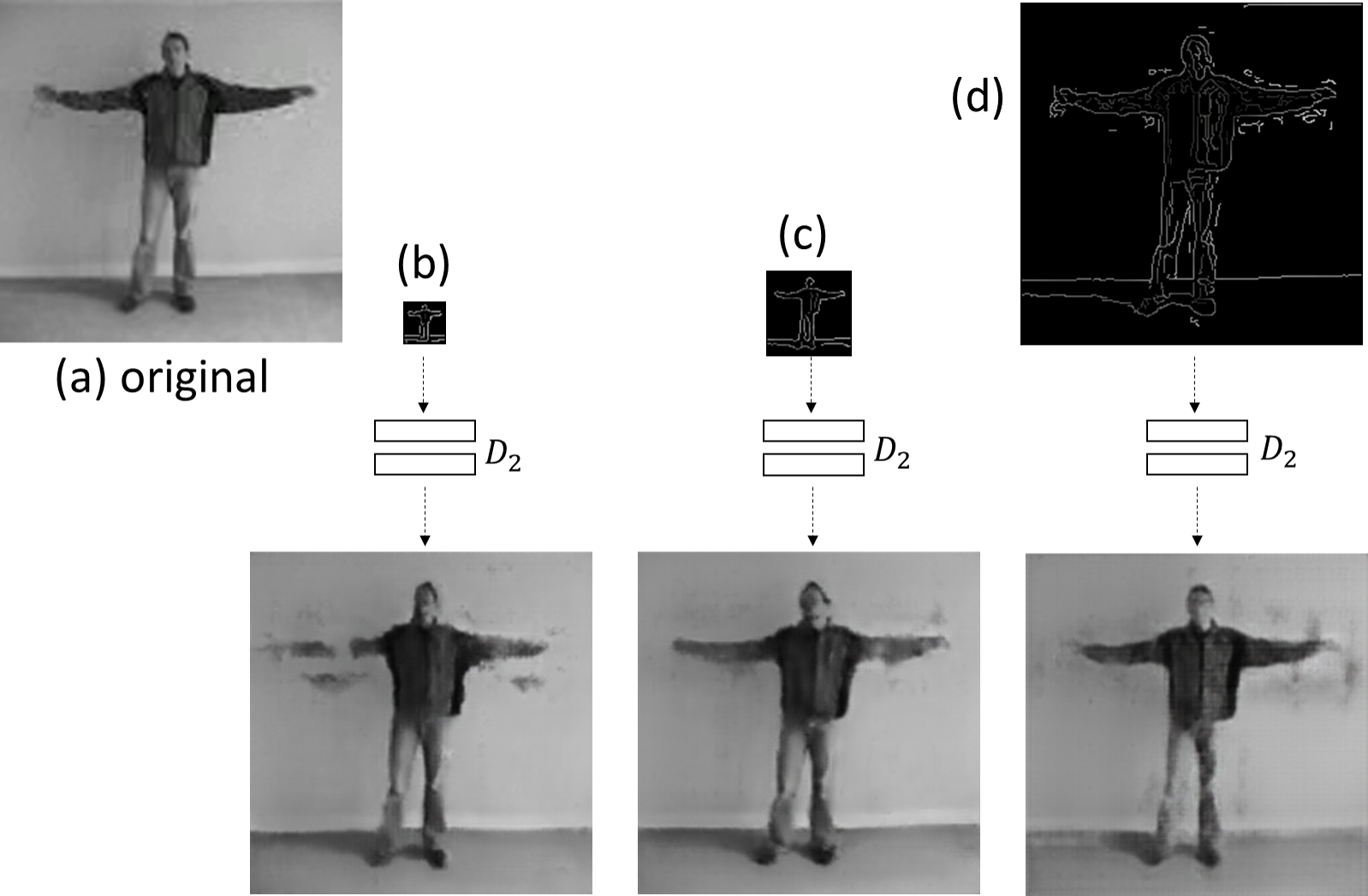}
\end{center}
   \caption{Performance of proposed framework against different downsampling levels: (a) original $256\times256$ frame, $X_G$. Reconstructions at scales (b) $32\times32$, (c) $64\times64$, (d) and $256\times256$. As the resolution increases, the reconstructed frames become more recognizable.}
\label{fig:downsampler_ex}
\end{figure}

To study the relative qualities of reconstructed frames using 
different levels of {\textit{downsampling}}, we employed three scaling sizes, $(1,1)$, $(4,4)$ and $(8,8)$. Hence, an original frame of size $256\times256$, became $256\times256$, $64\times64$, and $32\times32$, respectively (Figure~\ref{fig:downsampler_ex}). In addition, at the second stage of encoding we fixed \textit{soft edge detector} with quantization level $2$ ($Q_2$) without \textit{lossless compression}, to isolate the effects of downsampling. As the scaling size was increased, the reconstructed frames became more recognizable (Figure~\ref{fig:downsampler_ex}). Accordingly, the measured VQA scores also increased monotonously, as the size of $x_G$ was increased (Table~\ref{table:downsampler_QA}).

\subsection{Quality of reconstructed quantized frame}
\label{sec:exp_Q_edge}
\begin{figure}
\begin{center}
\includegraphics[scale=0.265]{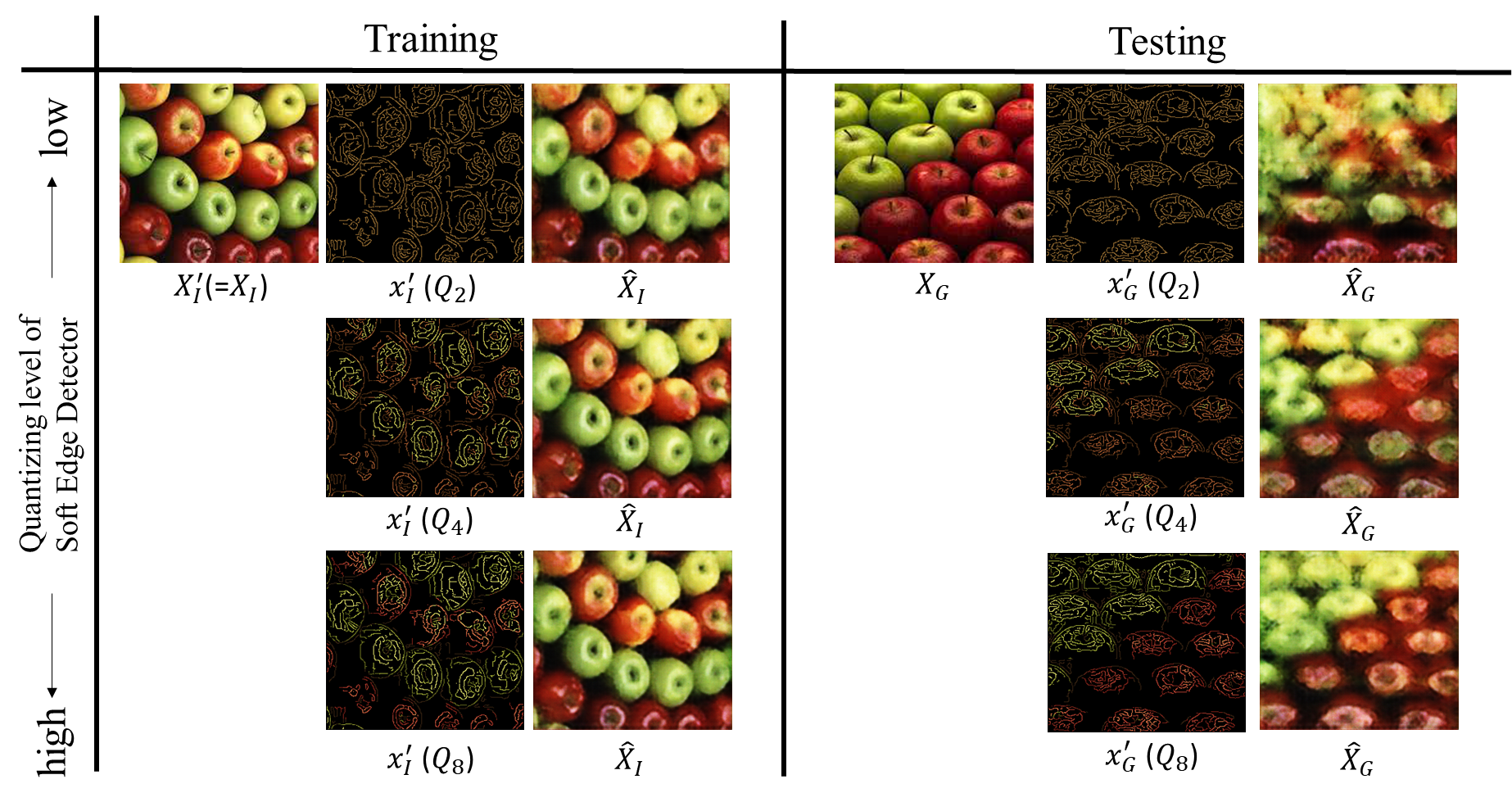}
\end{center}
   \caption{Performance of proposed framework against different quantization levels $k$ of the \textit{soft edge detector} ($Q_k$). As the quantization level is increased (more clusters), the reconstructed representations become more precisely and similar to an original frames}
\label{fig:soft_edge_detector}
\end{figure}

We also examined the relationship between the quality of reconstructed frames and the level of quantization delivered by the \textit{soft edge detector} (see Section~\ref{sec:Soft_edge_detector}). As a toy example, assume $E_1=E_2=I$, i.e, $X_I=X_I'$. $E_2$ taken to the \textit{soft edge detector} with quantization level $k$ ($Q_k$) without any \textit{downsampling} 
(Figure~\ref{figure:E_2}). First, we applied a quantization $k=2$ to generate $x_I'$. Then, $D_2$ was trained on a single pair ${X'_I}^{(1)}$ and ${x'_I}^{(1)}$ (both frames composed of $256 \times 256$ pixels). 
The trained $D_2$ was then applied on ${x'_G}^{(1)}$ to reconstruct ${\hat{X}_G}^{(1)}$ (Figure \ref{fig:soft_edge_detector}). 
We also repeated this experiment with $k=4$ and $k=8$. In this toy example, 
our network was trained on only one image, assuming no compression for $X_I$ ($X_I'=X_I$). 
In addition, at the second stage of encoding, no \textit{downsampler} or \textit{lossless compression} was used.

As the quantization level $k$ was increased, the quality of the reconstructed frames improved qualitatively ($\hat{X}_G$ in Figure \ref{fig:soft_edge_detector}). This improvement was also measured by quantitative VQA scores, which also indicated a trend of improvement with higher value of $k$ (Table \ref{table:soft_edge_detector_QA}).

\begin{table}
\begin{center}
\includegraphics[scale=0.4]{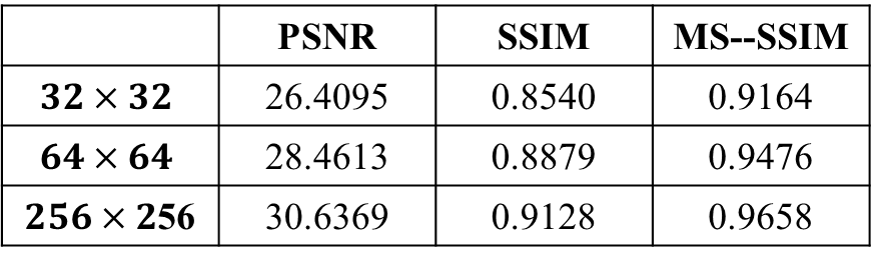}
\end{center}
   \caption{Video quality assessment of reconstructed frames in Figure~\ref{fig:downsampler_ex}. As the resolutions increased, the quality scores of the reconstructed frame increase monotonical.}
\label{table:downsampler_QA}
\end{table}

\subsection{Performance against key frame density}

In our framework, performance and bitrate are closely associated with $\alpha$, which we define as the ratio of the number of key frames to the total number of frames in one video. In our experiments, our scheme achieved fair performance using only a very small number of key frames, e.g., $\alpha=1\%$. In the experiments in the coming sections, we set $\alpha \leq 1.5\%$. 

\begin{table}
\begin{center}
\includegraphics[scale=0.4]{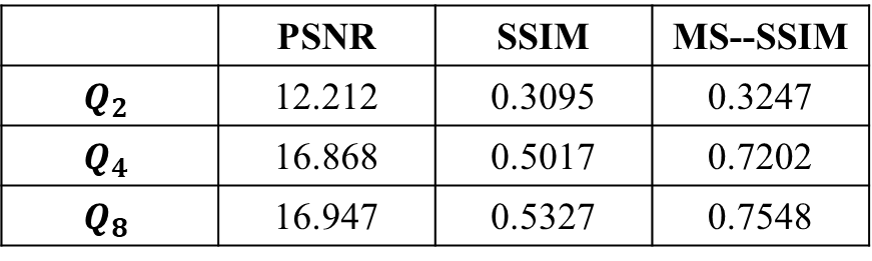}
\end{center}
   \caption{Video quality assessment of reconstructed frames in Figure~\ref{fig:soft_edge_detector}. As $k$ is increased, the quality of the reconstructed frames becomes improve.}
\label{table:soft_edge_detector_QA}
\end{table}

\subsection{Comparison with H.264}

One of our experimental results is summarized in Figure~\ref{fig:signiture}; seven consecutive frames from an original video and three compressed videos: (1) H.264 at 7.62 Kbps, (2) H.264 at 10.36 Kbps, and (3) our model at 7.14 Kbps, are depicted. Our model delivered noticeably better quality at the lowest bitrate than H.264. To train our network using a conditional GAN, 80 key frames ($N_I=80$) were collected from a single video having 8000 frames ($\alpha=1\%$), and assigned 2.02 Kbps of the overall 7.14 Kbps. Seven test frames were not included in the training set. We employed an $8 \times 8$ \textit{downsampler} and the $Q_8$ \textit{soft edge detector}. As a method of objective quantitative analysis, the scores produced by several leading perceptual video quality metrics (PSNR, mean SSIM, mean MS-SSIM, and VMAF scores) were plotted for the reconstructed videos. The red curves denotes the results of our model, while the blue curves are those for H.264. Our model was able to achieve much higher video quality scores below 10 Kbps than did H.264. Interestingly, our codec was able to successfully compress the video at less than 7.5 Kbps, while H.264 failed (no output).

\begin{figure}
\begin{center}
\includegraphics[scale=0.52]{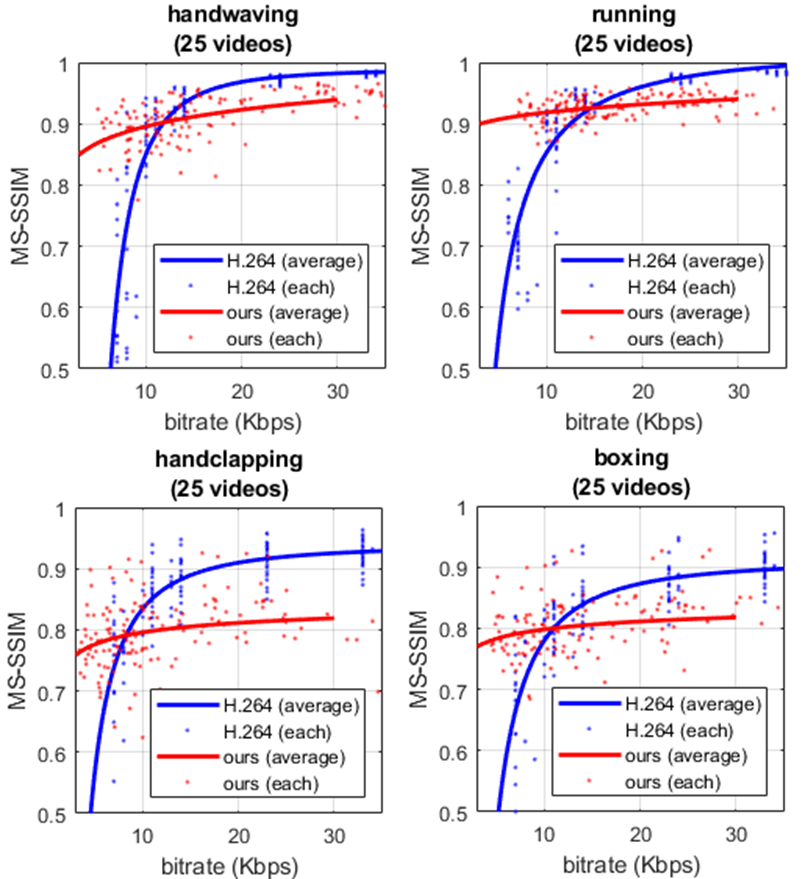}
\end{center}
   \caption{Rate-distortion curves (MS-SSIM) against bitrate for four semantic categories of 100 videos from the KTH dataset~\cite{schuldt2004recognizing}. The red curves and dots correspond to our model while the blue curves and dots correspond to H.264. In the very low bitrate region (below 10Kbps), our scheme yielded higher MS-SSIM scores. Similar results were observed on PSNR, SSIM and VMAF (see supplementary material).}
\label{fig:RD_100}
\end{figure}

To examine the performance of the proposed method, we implemented an experiment on videos from four semantic categories in the KTH ~\cite{schuldt2004recognizing}. Figure~\ref{fig:RD_100} plots RD curves of bitrate against VQA scores for 100 videos. The red curves correspond to our model while the blue curves represent the performance of H.264. Notably, in the below 10 Kbps, our scheme achieved significantly higher MS-SSIM scores than did H.264. Similar results were observed on PSNR, SSIM and VMAF (please see Supplementary material). 

Figure~\ref{fig:KTH_Youtube} shows the performance of our codec, as compared to H.264.  Figure~\ref{fig:KTH_Youtube}(a) shows selected frames from both an original KTH video and also compressed videos using H.264 at 9 Kbps, H.264 at 13 Kbps, and our compression model. The videos are aligned vertically for visual comparison. Our codec (4th row) delivered significantly better visual quality at low bitrates than did H.264. As a method of objective quantitative analysis, VQA metrics were plotted on the right side of Figure~\ref{fig:KTH_Youtube}. For example, our model achieved MS-SSIM of 0.9188 at 8.33 Kbps, while H.264 resulted in MS-SSIM score of 0.8884 at 13 Kbps. 

Since most of the videos in the KTH dataset contain simple, repeated object movements, we further validated our model on a variety of videos from the YouTube dataset~\cite{Charles16}, contain more diverse activities and impredictable movements. Figure~\ref{fig:KTH_Youtube}(b) visually compares compression performance on one YouTube video. Our proposed codec outperformed H.264 regards to perceptual quality versus bitrate. Similar results were observed on the other 30 videos from the YouTube pose dataset (see Supplementary material). 

\begin{figure*}
\begin{center}
\includegraphics[scale=0.36]{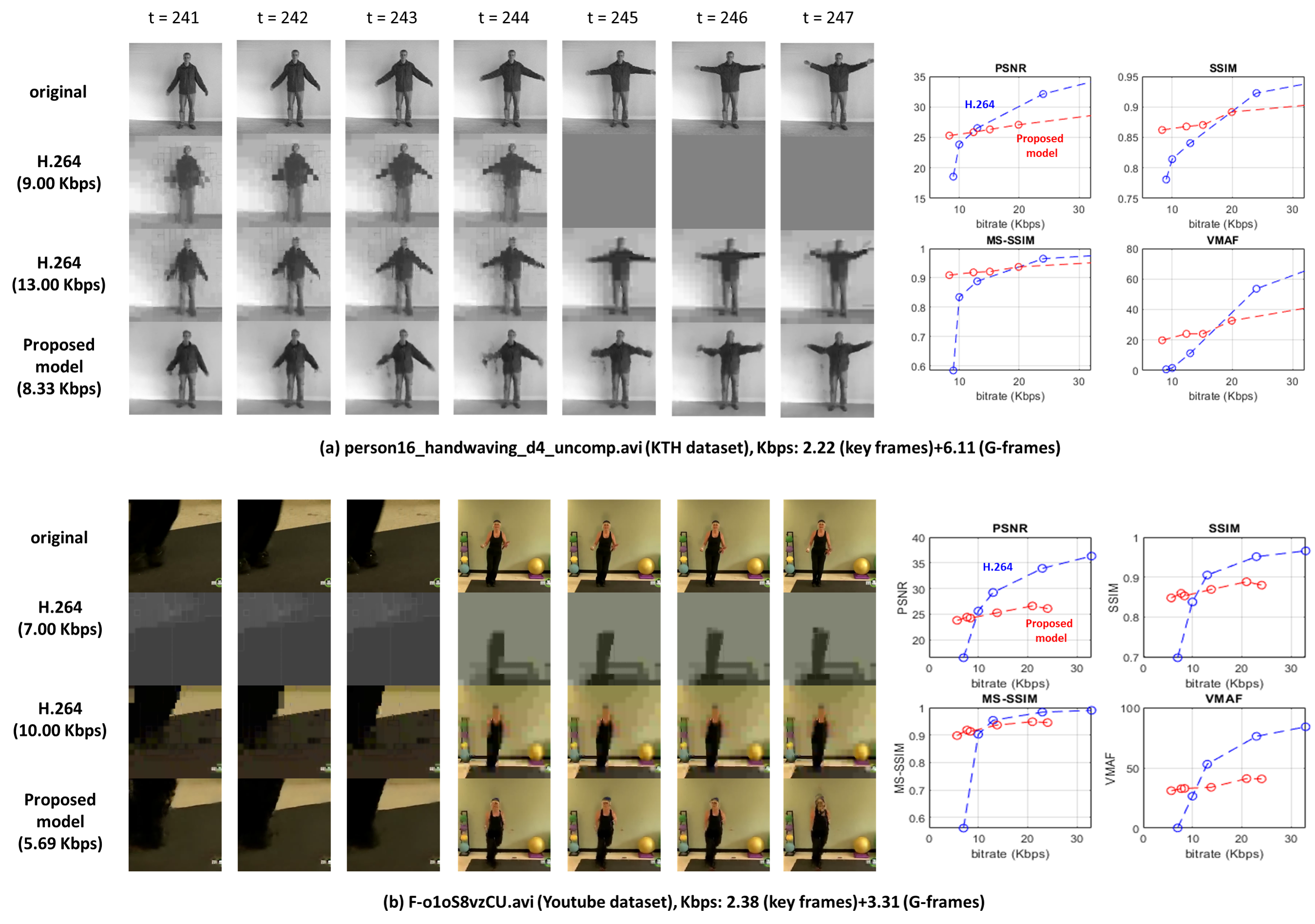}
\end{center}
   \caption{Two videos from (a) the KTH and (b) the YouTube dataset. Selected frames from original video and reconstructed videos using H.264 (low bitrate), H.264 (high bitrate), and the proposed model are aligned vertically along time. Our scheme demonstrated significantly better performance than the current standard codecs at low bitrates. 
   The scores produced by several leading perceptual video quality metrics were depicted on the right side. Please refer to the supplementary for reconstructed videos and results on additional 129 videos.}
\label{fig:KTH_Youtube}
\end{figure*}

\section{Limitations}

\begin{figure*}
\begin{center}
\includegraphics[scale=0.36]{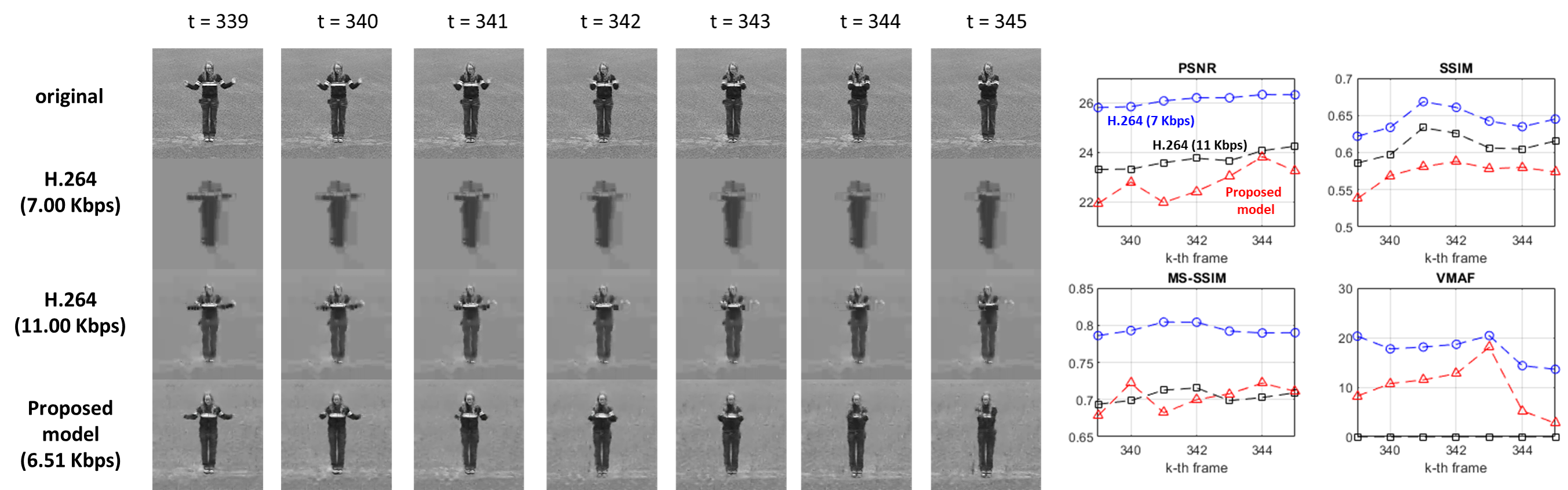}
\end{center}
   \caption{Example of lower VQA scores on videos compressed by our model than those compressed using H.264, despite the apparent better subjective quality produced by our model. VQA scores taken above 7 frames are depicted at the right side (please see \href{https://youtu.be/5pO-gm9R40g}{this video}).} \label{fig:VQA_limitation_4}
\end{figure*}

As it can be seen in Figure \ref{fig:RD_100}, our codec delivered worse VQA scores than H.264 at bitrates higher than 10 Kbps. This gap might be eliminated by using a larger (deeper) network than the very moderate one used here~\cite{wang2017high}. Moreover, we manually select $n(X_I)$ and the key frames although exhaustive search or a suboptimal Greedy algorithm could be used to select n(Xi) and the key frames. We also considered a few configurations of the \textit{downsampler} ($4\times4$ and $8\times8$) and the \textit{soft edge detector} ($Q_2$, $Q_8$ and $Q_{16}$). Beyond that, we could explore different choices of the edge detector, or the use of semantic labelling algorithms. 

The limitations of the current VQA metrics may also be relevant. We believe that most of this gap may be explained by a failure of modern VQA models to accurately assess the quality of a GAN-generated frame against a reference. It is quite possible that in this context, conventional VQA models may not accurately predict the subjective quality of GAN-compressed videos. As shown in Figure \ref{fig:VQA_limitation_4}, our scheme apparently reconstructs frames of better quality than H.264. For example, H.264 compressed at 7.00 Kbps resulted in unrecognizable reconstructed frames (many blurred blocks), while our model at 6.51 Kbps yielded more detailed frames than even H.264 at 11 Kbps. However, the VQA scores were often lower for our model than for H.264. For example, Figure \ref{fig:VQA_limitation_4}(b) plots the temporal evolution of objective video quality scores on seven consecutive frames. Remarkably, although the visual appearance of the video compressed by our model was much better than the H.264 result, the PSNR and SSIM scores at 6.51 Kbps (red lines) yielded a reverse relationship against H.264 at 7.00 and 11.00 Kbps (black and blue lines respectively). This is likely because the GAN-based compressor produces representations that appear very similar to the original, but differ numerically, e.g., over textured regions composed of similar, but differently arranged elements. This suggests that novel quality assessment metrics, designed for video based DNNs, are a worthy research area for further investigation.

Our video compression experiments are focused on very low bitrate ranges (3-30 Kbps) and encoding resolutions ($256 \times 256$). While these parameters are lower than traditionally used bitrate ladders, our goal in this paper was to demonstrate a proof of concept for our GAN-based model. 

\section{Conclusions}
We proposed a video compression framework that is based on conditional GANs guided by \textit{soft edge detection}. We showed that our scheme achieved better visual results and higher objective VQA scores than current standard video codecs at low bitrates. At higher bitrates, our model produces competitive visual results but worse VQA scores. Much of the reason for this is that the representations created by the GAN, while visually accurate, are not guided to be pixel-wise accurate, particularly on image textures. Our approach is re-establishing the importance of edges in modern DNN-based video compression architectures.  

In this work we have only relied on a moderately deep network structure. By training deeper networks, we believe that our framework will be able to obtain competitive performance against any scheme, whether standard, hybrid, or DNN-based, even at higher bitrates and/or resolutions.

{\small
\bibliographystyle{ieeetr}
\bibliography{egbib}
}
\end{document}